\documentclass[12pt]{article}
\usepackage{CJKutf8}
\usepackage{geometry}
\usepackage{float}
\usepackage{amsmath}
\usepackage{graphicx}
\usepackage{setspace}
\makeatletter
\usepackage{indentfirst}

\makeatother

\begin{document}
\begin{CJK}{UTF8}{bsmi}%
\pagenumbering{roman}

\title{Theoretical investigation of the bow shock location for the solar
wind interacting with the unmagnetized planet}

\author{I-Lin Yeh$^{1,2}$, Sunny W.Y. Tam$^{2}$, Po-Yu Chang$^{2}$}

\maketitle
$^{1}$Department of Physics, UC San Diego, USA

$^{2}$Institute of Space and Plasma Sciences, National Cheng Kung
University, Taiwan
\begin{abstract}
Martian bow shocks, the solar wind interacting with an unmagnetized
planet, are studied. We theoretically investigated how solar parameters,
such as the solar wind dynamic pressure and the solar extreme ultraviolet
(EUV) flux, influence the bow shock location, which is still currently
not well understood. We present the formula for the location of the
bow shock nose of the unmagnetized planet. The bow shock location,
the sum of the ionopause location and bow shock standoff distance,
is calculated in the gasdynamics approach. The ionopause location
is determined using thermal pressure continuity, i.e., the solar wind
thermal pressure equal to the ionospheric pressure, according to tangential
discontinuity. The analytical formula of the ionopause nose location
and the ionopause profile around the nose are obtained. The standoff
distance is calculated using the empirical model. Our derived formula
shows that the shock nose location is a function of the scale height
of ionosphere, the dynamic pressure of the solar wind and the peak
ionospheric pressure. The theoretical model implies that the shock
nose location is more sensitive to the solar EUV flux than solar wind
dynamics pressure. Further, we theoretically show that the bow shock
location is proportional to the solar wind dynamic pressure to the
power of negative C, where C is about the ratio of the ionospheric
scale height to the distance between bow shock nose and the planet
center. This theory matches the gasdynamics simulation and is consistent
with the spacecraft measurement result by Mars Express {[}Hall, et
al. (2016) J. Geophys. Res. Space Physics, 121, 11,474-11,494{]}. 

\end{abstract}
\pagebreak{}

\pagenumbering{arabic}

\section{Introduction}

In the nowadays space physics research, more research focuses on the
solar wind interacting with the magnetized planet than the unmagnetized
planet or weakly-magnetized planet. However, the study of Martian
bow shock is recently a hot topic due to the growing interests in
the exploration of Mars, a weakly-magnetized planet. Furthermore,
more and more data from the spacecraft measurement are available.
A theoretical model of the bow shock nose position has been derived
in this work.

\begin{figure}[H]
\begin{centering}
\includegraphics[width=0.9\columnwidth]{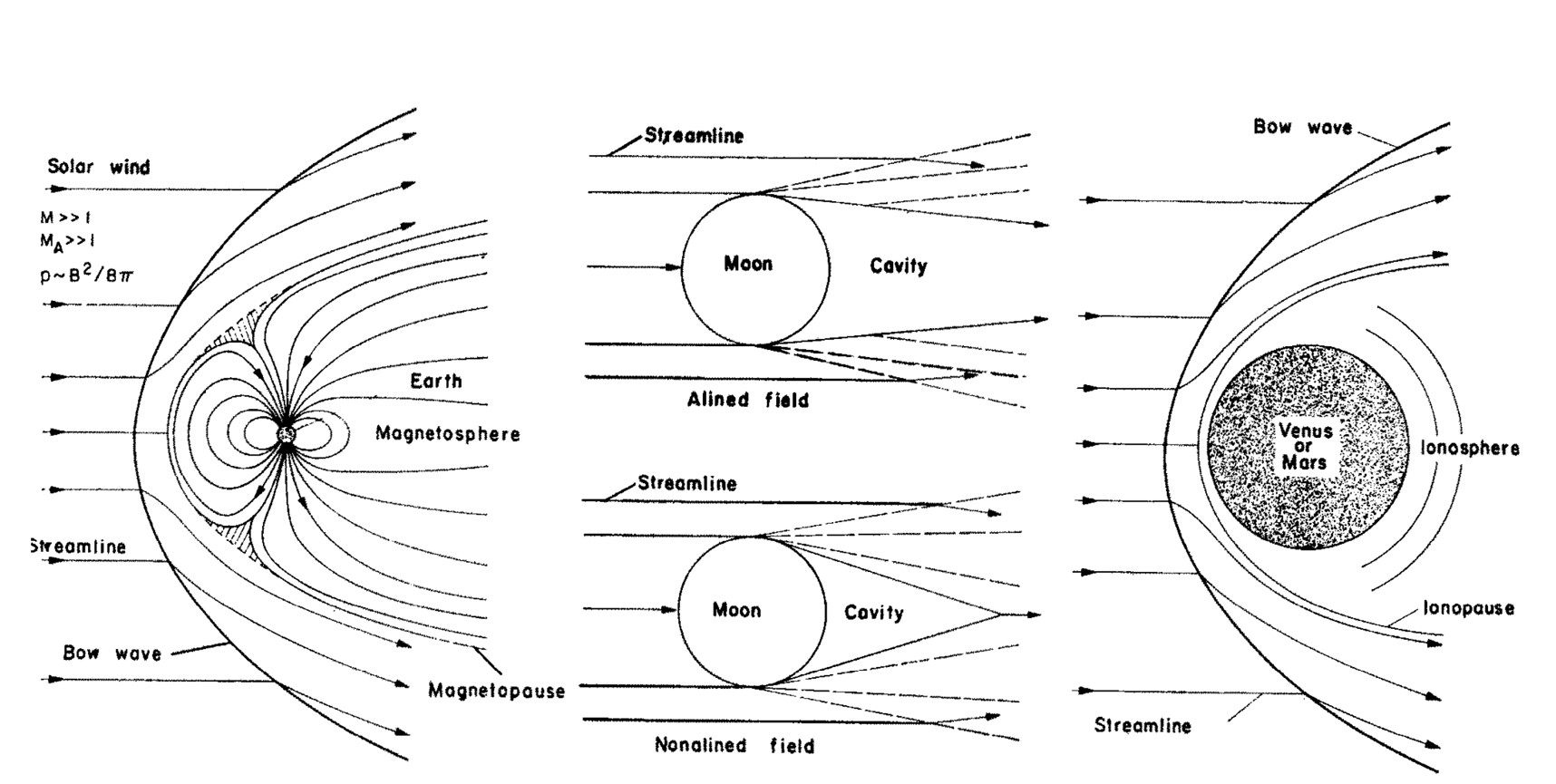}
\par\end{centering}
\caption{Schematic of solar wind past the (a) Earth (b) Moon (c) Mars and Venus.
Courtesy of Ref. \cite{spreiter1970solar}.\label{fig:Schematic-of-solar}}
\end{figure}

There are three types of interaction between the solar wind and an
obstacle in space: (1) solar wind interacting with the magnetized
obstacle like Earth. (2) solar wind interacting with the unmagnetized
obstacle with an atmosphere like Mars and Venus. (3) solar wind interacting
with the unmagnetized obstacle without atmosphere like moon. Figure
\ref{fig:Schematic-of-solar} is a schematic of these three types
of interaction.

The detached bow shock is formed because of the supersonic solar wind
and the deflection of the incident solar wind flow by the magnetosphere
or ionosphere. Since the moon has neither ionosphere nor magnetosphere,
no bow shock is formed in the solar wind interacting with the moon. 

\begin{figure}[H]
\begin{centering}
\includegraphics[width=0.7\columnwidth]{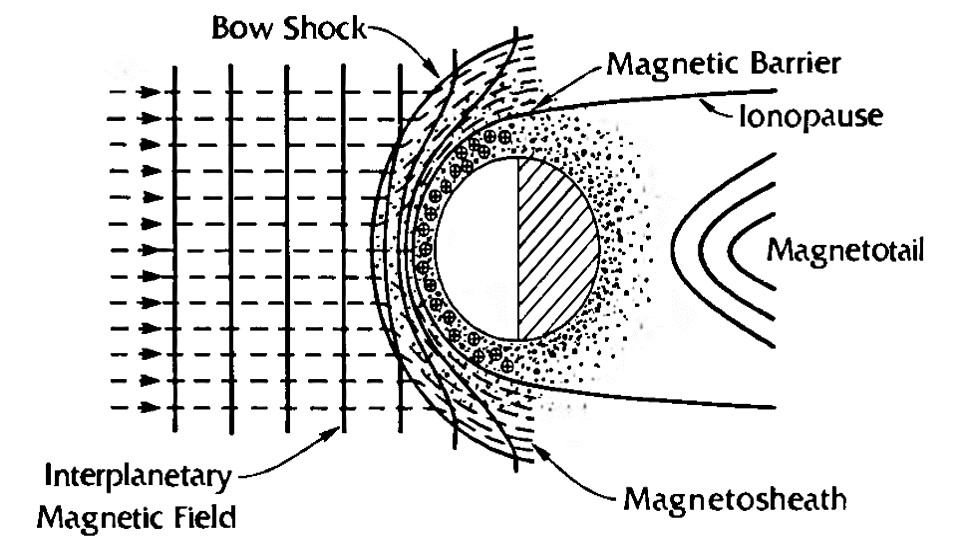}
\par\end{centering}
\caption{Schematic of the solar wind interaction with an unmagnetized planet
with an atmosphere. Courtesy of Ref.\cite{kivelson1995introduction}.
\label{fig:Schematic-unma} }
\end{figure}

Formation of the bow shock in plasma interaction with Mars, unmagnetized
planets with an atmosphere (Fig. \ref{fig:Schematic-unma}), is as
follows. First, the ionization by solar EUV radiation in the atmosphere
forms an ionospheric obstacle, acting as a conductor. The boundary
of the ionosphere is called ionopause. Then the solar-wind plasma
with its frozen-in field flows at a supersonic velocity toward the
conducting obstacle, resulting in the appearance of the bow shock. 

Martian bow shock has been detailedly studied by spacecraft measurement
and numerical simulation. Martian bow shock is formed by the interaction
between the solar wind and the Martian ionosphere. Recently, the first
measurement study\cite{vogt2015ionopause} of the ionopause from the
mission Mars Atmosphere and Volatile EvolutioN (MAVEN, 2014-present)
was released in 2015. This mission will provide us a deeper understanding
of the Martian bow shock. The shape of the bow shock is often modeled
using the least-squares fitting of an axisymmetric or non-axisymmetric
conic section\cite{farris1994determining,trotignon2006martian} with
the data from spacecraft measurement. On the other hand, the theoretical
model of the planetary bow shock location and shape can be seen in
the review paper by Spreiter\cite{spreiter1970solar,spreiter1966hydromagnetic,spreiter1995location},
Slavin\cite{slavin1981solar,slavin1983solar} and Verigin\cite{verigin2003planetary}.
An efficient computational model for determining the global properties
of the solar wind past a planet based on axisymmetric magnetohydrodynamics
was proposed by Spreiter\cite{spreiter1980new}. The specific study
of the magnetohydrodynamics simulation for the solar wind interaction
with Mars can be seen in Ref.\cite{spreiter1992computer} and Ref.\cite{ma2002three}. 

\begin{figure}[H]
\begin{centering}
\includegraphics[width=0.9\columnwidth]{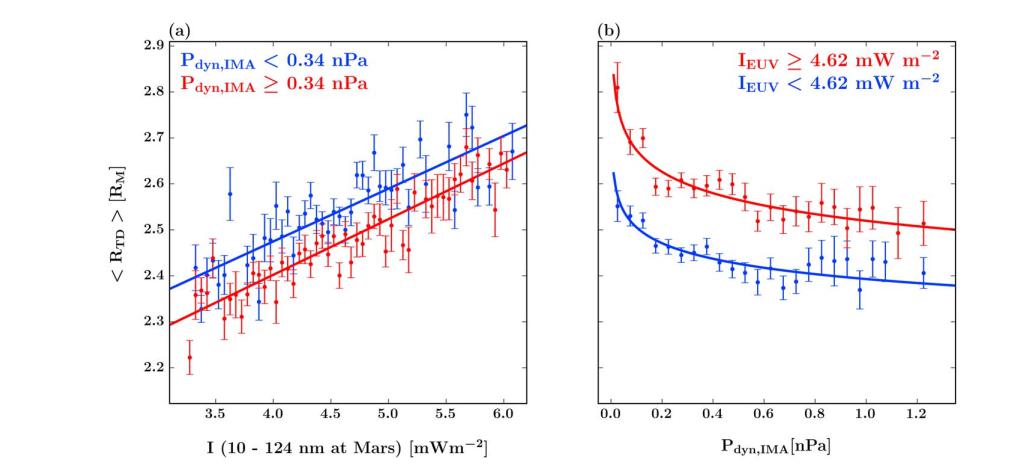}
\par\end{centering}
\caption{The response of the location of the Martian bow shock with the solar
parameters. (a) Bow shock location against solar EUV radiance. (b)
Bow shock location against solar wind dynamic pressure. Courtesy of
\cite{hall2016annual}.\label{fig:The-response-of}}
\end{figure}

However, how the factors influence the location of the Martian bow
shock is not well understood. The main factors impacting the bow shock
position are the solar wind dynamic pressure $P_{dyn}=\rho_{\infty}v_{\infty}^{2}$
and solar EUV flux $l_{euv}$. According to the fitting results (Fig.
\ref{fig:The-response-of}) from the data of Mars Express Analyser
of Space Plasma and EneRgetic Atoms (ASPERA-3)\cite{hall2016annual},
it is shown that the bow shock location ($r_{s}$) reduces in altitude
with increasing solar wind dynamic pressure in the relation $r_{s}\propto P_{dyn}^{-0.02}$
and increases in altitude with increasing solar EUV flux in the relation
$r_{s}\propto0.11\,l_{EUV}$. It means that the bow shock position
is more sensitive to the solar EUV flux than the solar wind dynamic
pressure.

Other parameters controlling the Martian bow shock location are the
intense localized Martian crustal magnetic fields\cite{vignes2002factors},
the magnetosonic Mach number\cite{edberg2010magnetosonic}, the interplanetary
magnetic fields and the convective electric field\cite{edberg2008statistical}.
In this thesis, we will mainly focus on the dependence of the solar
wind dynamics pressure and the ionospheric pressure, which is dependent
on the solar EUV radiation.

It is not well understood how the location of the Martian bow shock
is influenced by the solar parameters such as solar wind dynamic pressure
and EUV flux. In this thesis work, we study the location of the bow
shock generated from the interaction between the solar wind and unmagnetized
planet in the theoretical aspect. The formula for the shock nose location
as a function of the solar wind dynamic pressure and EUV flux will
be presented and compared with the spacecraft measurement. Past studies
by others are all related to observation, but this study provides
a theory to explain the spacecraft measurement results. On the other
hand, we are not going to study fine structure in the transition region
of a shock and the shock microphysics, even though it is more theoretically
fascinating. The dissipation mechanism for the shock will not influence
the location of the bow shock, so we use the ideal hydrodynamics formulation
throughout this work. Note that the thickness of the discontinuity
surface is zero under the ideal hydrodynamics description. 

The analytical theory is essential because the relationship between
each physical quantities can be directly known in the formula. However,
in simulation, to know the results from different conditions requires
different runs, which is very numerically intensive especially for
multi-scale and multi-physics simulation. This work will be beneficial
for both space physics and laboratory astrophysics research. In laboratory
astrophysics research. 

In this paper, we will focus on theoretically determining the location
of the nose of the bow shock for the solar wind past the unmagnetized
planet. The formula for shock nose location is developed and compared
to the results from the hydrodynamics simulation and spacecraft measurement.
In section \ref{sec:Determination-of-the}, we will give a detailed
derivation of the formula for the shock nose position. The comparison
of our formula and the spacecraft measurement results will be shown
in section \ref{subsec:Comparison-theory}. In section \ref{sec:Conclusion},
the conclusion of the thesis will be given. 

\pagebreak{}

\section{Determination of the location of the bow shock nose \label{sec:Determination-of-the}}

We theoretically investigate the bow shock location as a function
of the solar wind and the ionospheric conditions, such as solar wind
dynamic pressure $\rho_{\infty}v_{\infty}^{2}$, ionospheric scale
length $H$, ionospheric peak pressure $P_{M,i}$ and the location
of the ionospheric peak pressure $r_{M,i}$. The equation of bow shock
location is derived and will be used to design the future experiments.
We only focus on the nose location of the bow shock but not the whole
shape profile of the bow shock. The shape of the bow shock and the
ionopause is assumed to be symmetric around the x-axis. 

\begin{figure}[H]
\begin{centering}
\includegraphics[width=0.7\columnwidth]{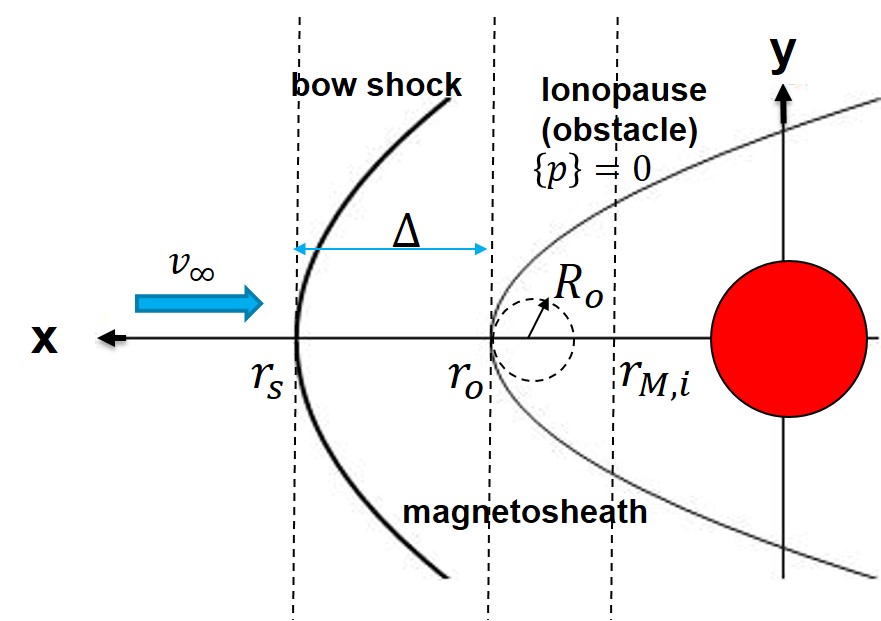}
\par\end{centering}
\caption{Definition of each variable used in the report \label{fig:Definition-of-varaibles}}
\end{figure}

The schematic of the bow shock and the obstacle boundary (ionopause)
is shown in the Fig.  \ref{fig:Definition-of-varaibles}. We use the
following symbols to represent the geophysical quantities in the report:
$r_{o}$ is nose positions of the obstacle, $r_{s}$ is nose positions
of the bow shock, $r_{M,i}$ is the location inside ionosphere where
maximal thermal pressure occurs, $\Delta$ is the bow shock standoff
distance, i.e., the distance between ionopause nose position and bow
shock nose position. 

The goal is to determine the shock nose location $r_{s}$:
\begin{equation}
r_{s}=r_{o}+\Delta.\label{eq:rs=00003Dro+delta}
\end{equation}
Ionopause nose location $r_{o}$ is calculated using the continuity
of the thermal pressure by tangential discontinuity\cite{spreiter1970solar,spreiter1966hydromagnetic,verigin2003planetary};
bow shock standoff distance is calculated by the empirical formula\cite{farris1994determining,verigin2003planetary}.

We first introduce the hydrodynamics boundary conditions in subsection
\ref{subsec:Hydrodynamics-boundary-condition}. The derivation of
the ionopause nose location and the radius of curvature at ionopause
nose are in subsection \ref{subsec:Ionopause-(obstacle-boundary)}.
The standoff distance formula is introduced in subsection \ref{subsec:Bow-shock-standoff}.
Finally, the formula of the bow shock nose location is shown in subsection
\ref{subsec:shock location}. The comparison of the theory and the
observation results will be given in next section (section \ref{subsec:Comparison-theory}).

\subsection{Hydrodynamics boundary condition\label{subsec:Hydrodynamics-boundary-condition}}

Hydrodynamics formulation is used throughout this thesis. Ideal magnetohydrodynamics
equations are
\begin{align}
\frac{\partial\rho}{\partial t}+\nabla\cdot\rho\,\vec{v} & =0,\nonumber \\
\rho(\frac{\partial\vec{v}}{\partial t}+\vec{v}\cdot\nabla v) & =-\nabla P+\frac{1}{\mu_{0}}(\nabla\times\vec{B})\times\vec{B},\nonumber \\
\frac{\partial\vec{B}}{\partial t} & =\nabla\times\vec{v}\times\vec{B},\nonumber \\
\frac{\partial P}{\partial t}+\vec{v}\cdot\nabla P & =-\gamma\,P\:\nabla\cdot\vec{v},\label{eq:Ideal_mhd}
\end{align}
where $p$ is the pressure, $\rho$ is the mass density, $\vec{v}$
is the velocity and $\vec{B}$ is the magnetic field. The first one
is the continuity equation, the second is the momentum equation, the
third is Faraday's law and the last is the entropy conservation equation,
or adiabatic equation. Here we assume the gas follows the polytropic
condition and adiabatic process. 

Magnetohydrodynamics equations can be reduced to hydrodynamics equations
under the condition that the magnetic pressure term is much smaller
than the thermal pressure term in the right-hand side of the momentum
equation (second equation in Eq. \ref{eq:Ideal_mhd})
\begin{equation}
|\frac{\frac{1}{\mu_{0}}(\nabla\times\vec{B})\times\vec{B}}{\nabla P}|\approx\frac{B^{2}/2\mu_{0}}{P}=1/\beta,\label{eq:beta}
\end{equation}
where plasma beta $\beta$ is defined as thermal pressure divided
by magnetic pressure. For the condition of solar wind past the unmagnetized
planet, $\beta$ is much larger than 1 in both space and laboratory,
so we can neglect the force term containing the magnetic field in
the momentum equation, reducing the magnetohydrodynamics formulation
to pure hydrodynamics formulation
\begin{align}
\frac{\partial\rho}{\partial t}+\nabla\cdot\rho\,\vec{v} & =0,\nonumber \\
\rho(\frac{\partial\vec{v}}{\partial t}+\vec{v}\cdot\nabla v) & =-\nabla p,\nonumber \\
\frac{\partial P}{\partial t}+\vec{v}\cdot\nabla P & =-\gamma\,P\:\nabla\cdot\vec{v.}\label{eq:hydro_equ}
\end{align}
Throughout the paper, we will use hydrodynamics formulation instead
of magnetohydrodynamics due to the high beta condition. The first
one is the mass conservation equation, the second is the momentum
equation for ideal fluid, or Euler equation and the third is the adiabatic
equation.

The boundary condition for steady-state ideal hydrodynamics\cite{landau1987fluid}
is 
\begin{align}
\left[\rho v_{n}\right] & =0\nonumber \\
\left[P+\rho v_{n}^{2}\right] & =0\nonumber \\
\left[\rho v_{n}\vec{v_{t}}\right] & =0\nonumber \\
\left[v_{n}(\frac{\rho v^{2}}{2}+\frac{\gamma\,P}{\gamma-1})\right] & =0,\label{eq:hydro_BC}
\end{align}
where the subscript $n$ and $t$ are the normal direction and tangential
direction, respectively. The brackets mean the difference of the quantity
between both sides of the boundary. The Eq. \ref{eq:hydro_BC} indicates
the continuity of the mass flux, momentum flux and energy flux. Note
that the discontinuity surfaces of the ionopause and the bow shock
are zero thickness under the description of the dissipationless ideal
(magneto)hydrodynamics\cite{landau1987fluid,zel2012physics}.

In our study, we are interested in two types of boundary: 
\begin{itemize}
\item Tangential discontinuity\cite{spreiter1970solar,spreiter1966hydromagnetic,landau1987fluid}
at the ionopause
\begin{align}
\left[\rho\right]\neq & 0\nonumber \\
\left[\vec{v_{t}}\right]\neq & 0\nonumber \\
v_{n}= & 0\nonumber \\
\left[P\right]= & 0.\label{eq:tangential discontinuity}
\end{align}
The normal velocity is zero in the tangential discontinuity. We will
utilize the continuity of the thermal pressure to determine the location
and the radius of curvature at the ionopause nose. Furthermore, we
can observe that there is a density jump across the ionopause according
to tangential discontinuity.
\item Shock waves\cite{spreiter1970solar,spreiter1966hydromagnetic,landau1987fluid}
at the bow shock front
\begin{align}
\left[\rho v_{n}\right]= & 0\nonumber \\
\left[\vec{v_{t}}\right]= & 0\nonumber \\
\left[P+\rho v_{n}^{2}\right]= & 0\nonumber \\
\left[\frac{v_{n}^{2}}{2}+\frac{\gamma\,P/\rho}{\gamma-1}\right]= & 0.\label{eq:shock}
\end{align}
For our purpose of the study of the global phenomenon like bow shock
position, the ideal fluid description is enough. We are not going
to study microphysics such as the shock formation mechanism, so the
dissipation process of the shock will not be discussed throughout
the thesis. In general, the shock in the space is formed in a collisionless
magnetized environment and the shock dissipation mechanism is the
wave-particle interaction. 
\end{itemize}

\subsection{Overview of the theory}

\begin{figure}[H]
\begin{centering}
\includegraphics[width=0.7\columnwidth]{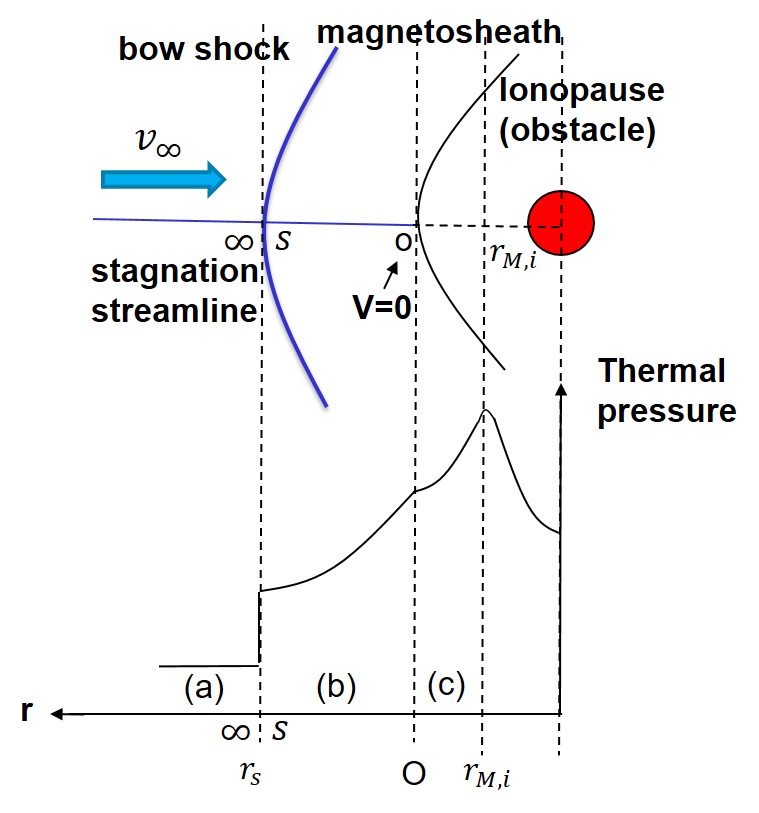}
\par\end{centering}
\caption{Thermal pressure along the stagnation streamline. \label{fig:Thermal-pressure-along} }
\end{figure}

In this subsection, we have an overview of all the theories which
are used for calculating the location of the bow shock nose in terms
of the solar wind and the ionospheric conditions. Fig. \ref{fig:Thermal-pressure-along}
shows the variation of the thermal pressure along the stagnation streamline.

\begin{itemize}
\item (a) Solar wind
\begin{itemize}
\item In the region of the solar wind, the thermal pressure can be expressed
as a function of the dynamic pressure and the sonic Mach number, i.e.,
$P_{\infty}=\rho_{\infty}v_{\infty}^{2}/(M_{\infty}^{2}\,\gamma)$.
\end{itemize}
\item (a) – (b) Bow shock
\begin{itemize}
\item At the bow shock, momentum flux conservation in the normal shock relation
is used 

\begin{equation}
P_{\infty}+\rho_{\infty}v_{\infty}^{2}=P_{s}+\rho_{s}v_{s}^{2}.\label{eq:pcons}
\end{equation}
Note that the entropy increases across the shock.
\end{itemize}
\item (b) Magnetosheath
\begin{itemize}
\item Within the magnetosheath, the plasma follows the process of the isentropic
compression, i.e., the combination of the energy conservation of the
compressible flow (Bernoulli equation)

\begin{equation}
\frac{2\gamma}{\gamma-1}P+\rho\,v^{2}={\rm constant},\label{eq:ber}
\end{equation}
and the adiabatic relation

\begin{equation}
P\,V^{\gamma}={\rm constant,}\label{eq:adia}
\end{equation}
where $V$ is the volume. We can observe that the sum of the thermal
pressure and the dynamic pressure is conserved before and after shock,
but not conserved along the stagnation streamline within the magnetosheath.
Therefore, the value of $P+\rho v^{2}$ before shock is not the same
as that at stagnation point. 
\end{itemize}
\item (b)-(c) Ionopause
\begin{itemize}
\item At the ionopause, the thermal pressure is continuous according to
tangential discontinuity.
\end{itemize}
\item (c) Upper ionosphere
\begin{itemize}
\item Within the upper ionosphere, the hydrostatic equilibrium (the balance
between the gravity force and the pressure gradient) is assumed.
\end{itemize}
\end{itemize}

\subsection{Ionopause (obstacle boundary) \label{subsec:Ionopause-(obstacle-boundary)}}

Ionopause, the boundary of the ionosphere, is the location of the
thermal pressure balance according to tangential discontinuity.\cite{spreiter1970solar,verigin2003planetary}.
We first investigate the pressure variation on the center line, then
the ionopause nose location $r_{o}$ and the radii of curvature at
the ionopause nose $R_{o}$.

\subsubsection{Thermal pressure at the ionopause\label{subsec:Thermal-pressure-ionospause}}

Ionopause profile is determined by the thermal pressure continuity
at both sides of the ionopause according to tangential discontinuity.
Here we discuss the thermal pressure at both sides of the ionopause
respectively. 
\begin{itemize}
\item Thermal pressure at the inner side of the ionopause 

The thermal pressure in the ionosphere is assumed to be spherical
symmetric and at hydrostatic equilibrium\cite{spreiter1970solar}
in equivalence to the balance between pressure gradient and gravity
force. So the thermal pressure inside the ionosphere can be expressed
as
\begin{equation}
P_{i}(r)=P_{M,i}\ {\rm exp}(\frac{r_{M,i}-r}{H}),\label{eq:exp of pion-1}
\end{equation}
where $P_{i}(r)$ is the pressure inside the ionosphere, $r_{M,i}$
is the location inside ionosphere where peak thermal pressure $P_{M,i}$
occurs and $H=k_{B}T/mg$ is the scale height in which $m=1.67\times10^{-24}g$
is the mass for a singly ionized hydrogen, $k_{B}$ is Boltzmann's
constant and $T$ is the absolute temperature for plasma and assumed
to be constant inside the ionosphere.
\item Thermal pressure at the outer side of the ionopause

We use Rayleigh pitot tube formula\cite{spreiter1966hydromagnetic,spreiter1992computer,landau1987fluid}
to obtain the thermal pressure just outside the ionosphere as a function
of the solar wind dynamic pressure. Rayleigh pitot tube formula is
used for the stagnation pressure at the blunt body nose with a detached
bow shock. It is derived in two steps: (1) applying the hydrodynamic
normal shock jump condition to get the downstream thermal pressure;
(2) applying the isentropic compression to determine the thermal pressure
at the stagnation point with Bernoulli's law on the stagnation streamline
within the magnetosheath. The rigorous derivation is shown in Appendix.
The Rayleigh pitot tube formula is given as
\begin{equation}
P_{o}=P_{\infty}M_{\infty}^{2}(\frac{\gamma+1}{2})^{(\gamma+1)/(\gamma-1)}\frac{1}{\left[\gamma-(\gamma-1)/(2M_{\infty}^{2})\right]^{1/(\gamma-1)}},\label{eq:Rayleigh}
\end{equation}
 where $P_{o}$ is the thermal pressure at the ionopause nose, $P_{\infty}$
is the thermal pressure of the solar wind, $M_{\infty}$ is the sonic
Mach number of the solar wind and $\gamma$ is the specific heat ratio.
Then, we plug $M_{\infty}=\frac{v_{\infty}}{\sqrt{\gamma p_{\infty}/\rho_{\infty}}}$
into the Rayleigh pitot tube formula, the relationship between thermal
pressure at the ionopause $P_{o}$ as a function of solar wind dynamic
pressure $\rho_{\infty}v_{\infty}^{2}$ can be expressed as
\begin{equation}
P_{o}=k\rho_{\infty}v_{\infty}^{2},\label{eq:ionos pres}
\end{equation}
 where 
\begin{equation}
k=(\frac{\gamma+1}{2})^{(\gamma+1)/(\gamma-1)}\frac{1}{\gamma\left[\gamma-(\gamma-1)/(2M_{\infty}^{2})\right]^{1/(\gamma-1)}}.\label{eq:k}
\end{equation}
 For $\gamma=5/3$ and $M_{\infty}\gg1$, this relation can be simplified
to $k=0.88$.
\end{itemize}

\subsubsection{Nose position of the ionopause $r_{o}$}

\begin{figure}[H]
\begin{centering}
\includegraphics[width=0.7\columnwidth]{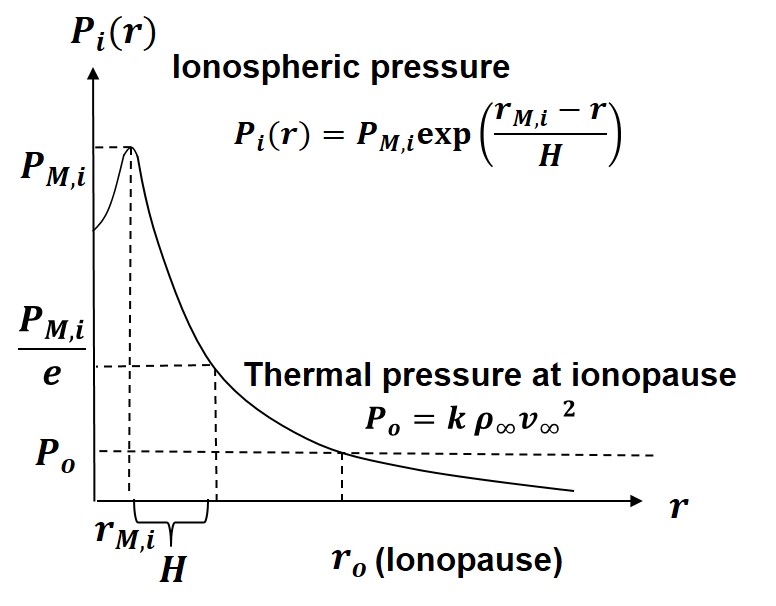}
\par\end{centering}
\caption{The schematic of the ionopause, where the ionsperic pressure is equal
to solar wind thermal pressure.\label{fig:pressure balance}}
\end{figure}

The formula of the ionopause nose position $r_{o}$ is determined
by the thermal pressure continuity ( Fig. \ref{fig:pressure balance})
at the ionopause according to tangential discontinuity:

\begin{equation}
P_{o}=P_{i}(r_{o}).\label{eq:pi=00003Dpinf}
\end{equation}

By solving Eq. \ref{eq:pi=00003Dpinf} with the expression of the
thermal pressure at the both side of the ionopause (Eq. \ref{eq:exp of pion-1}
and Eq. \ref{eq:ionos pres}), the formula of the nose position of
the bow shock $r_{o}$ can be derived as 
\begin{equation}
r_{o}=r_{M,i}+H\ {\rm ln}(\frac{P_{M,i}}{k\rho_{\infty}v_{\infty}^{2}}).\label{eq:exp of ro}
\end{equation}

The derived equation of the nose position (Eq. \ref{eq:exp of ro})
of the ionopause is reasonable: the shorter the scale height $H$
or the larger the dynamic pressure of the solar wind $\rho_{\infty}v_{\infty}^{2}$,
then ionopause closer to planet surface.

\subsubsection{Radius of curvature at ionopause nose $R_{o}$\label{subsec:Radius-of-curvature}}

In this section, we analytically calculate the radius of curvature
at ionopause nose $R_{o}$ by solving the ionopause profile equation
near the ionopause nose. Since the ionopause is symmetric at x-axis,
the ionopause profile can be expressed as $x=x(y)$ . We do the Taylor
expansion at $y=0$ of the ionopause profile $x=x(y)$, then we can
get the equation of the ionopause profile at the vicinity of the ionopause
nose 
\begin{equation}
x(y)=x(0)+(y-0)\,x'(0)+\frac{1}{2}(y-0)^{2}x"(0)+...\label{eq:expansion}
\end{equation}
Note that on the ionopause profile, $x(0)=r_{o}$ and $x'(0)=0$.
Furthermore, by the definition of the radius of curvature $R(y)=|\frac{(1+x'(y))^{3/2}}{x"(y)}|$,
the radius of curvature at ionopause nose ($y=0$) can be written
as $R_{o}=-1/x"(0)$. Thus, the equation of the ionopause profile
near the ionopause nose can be reduced to a quadratic equation
\begin{equation}
x=r_{o}-\frac{1}{2R_{o}}y^{2}.\label{eq:expansion with Ro}
\end{equation}
Here we neglect the third and higher order term of the Taylor expansion. 

\begin{figure}[H]
\begin{centering}
\includegraphics[width=0.7\columnwidth]{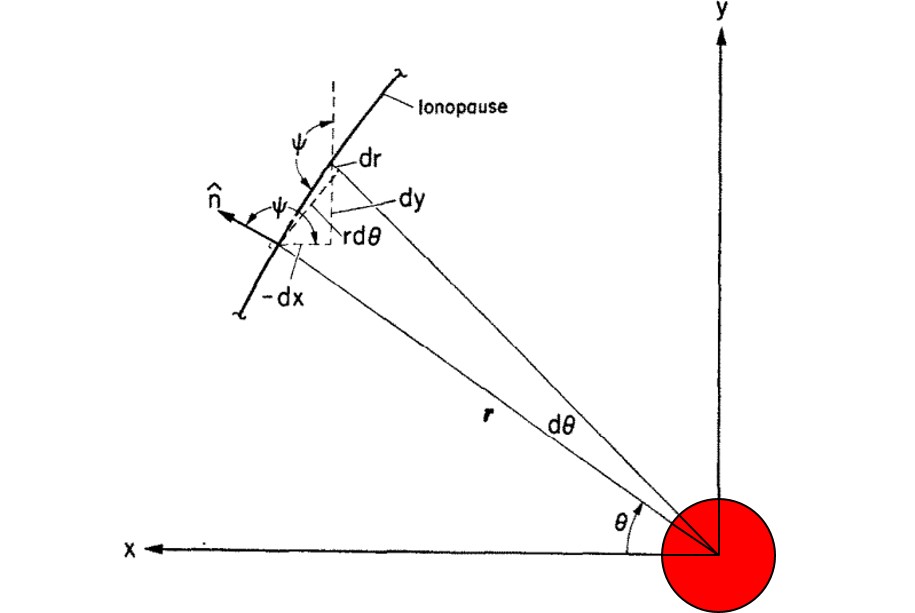}
\par\end{centering}
\caption{Element of the ionopause and the coordinate. Modified figure from
Ref. \cite{spreiter1970solar}. \label{fig:psi} }
\end{figure}

The whole ionopause profile can be determined by the thermal pressure
continuity at the ionopause according to tangential discontinuity,
that is, the thermal pressure is equal at the outer side and the inner
side of the ionopause. 
\begin{equation}
k\rho_{\infty}v_{\infty}^{2}\,{\rm cos^{2}}\psi=P_{i}(r),\label{eq:therbal}
\end{equation}
where $\psi$ is the angle between $v_{\infty}$ and the normal to
ionopause, which is shown in Fig. \ref{fig:psi}. 

 The left hand side of the equation is the thermal pressure approximated
at the outer side of the ionopause deviated from the nose position
and the right hand side is the ionospheric pressure we introduced
in section \ref{subsec:Thermal-pressure-ionospause}. Since $k\rho_{\infty}v_{\infty}^{2}$
is the ionospheric pressure at the ionopause nose $P_{i}(r_{o})$,
the equation of the ionopause profile (Eq.\ref{eq:therbal}) can be
reduced to

\begin{equation}
P_{i}(r_{o})\,{\rm cos^{2}}\psi=P_{i}(r).\label{eq:preba}
\end{equation}
By the geometric relation, $cos^{2}\psi$ can be expressed as
\begin{equation}
cos^{2}\psi=\left(\frac{dy}{ds}\right)^{2}=\frac{\left(r\,d\theta\:{\rm cos}\theta+dr\,{\rm sin}\theta\right)^{2}}{dr^{2}+(r\,d\theta)^{2}}.\label{eq: cospsi}
\end{equation}
We substitute the cosine relation in Eq.\ref{eq: cospsi} into the
pressure continuity equation at the ionopause (Eq.\ref{eq:preba}),
we get
\begin{equation}
P_{i}(r_{o})\frac{\left({\rm cos}\theta+(\frac{dr}{r\,d\theta})\,{\rm sin}\theta\right)^{2}}{\left(\frac{dr}{r\,d\theta}\right)^{2}+1}=P_{i}(r).\label{eq: pres_balance_polar}
\end{equation}
Then we solve for $dr/r\:d\theta$ to obtain 
\begin{equation}
\frac{dr}{r\,d\theta}=\frac{-P_{i}(r_{o})\,{\rm sin}2\theta+2\sqrt{P_{i}(r)P_{i}(r_{o})-P_{i}^{2}(r)}}{2(P_{i}(r_{o}){\rm sin^{2}}\theta-P_{i}(r))}.\label{eq: dr/rdtheta}
\end{equation}
This is the differential equation for the ionopause profile, which
can be solved numerically\cite{spreiter1970solar} with the initial
condition $r\left(\theta=0\right)=r_{o}$. Note that the ionopause
is symmetric about the $x=0$ axis ($\theta=0$), so the first-order
derivative at the ionopause nose is zero, that is, 
\begin{equation}
\frac{1}{r(0)}\frac{dr}{d\theta}(0)=\frac{1}{r_{o}}\frac{dr}{d\theta}(0)=0.\label{eq:derir0}
\end{equation}
In the second term of the numerator in the ionopause profile differential
equation (Eq.\ref{eq: dr/rdtheta}), it contains a square root. The
value of the quantity inside the square root must be equal or larger
than zero, or the square root term will become imaginary, which is
physically unallowable. So, we can get 
\begin{equation}
P_{i}(r)P_{i}(r_{o})-P_{i}^{2}(r)\geq0,\label{eq:ppo}
\end{equation}
where $P_{i}(r)$ is the ionospheric pressure exponentially decaying
outward because of hydrostatic equilibrium. Then we can obtain that
the ionopause profile must follow the condition 
\begin{equation}
r\geq r_{o}.\label{eq:rlargerro}
\end{equation}
The equality occurs at the ionopause nose.

For our purpose of deriving the radius of curvature at the ionopause
nose, we only have to focus on the vicinity of the ionopause nose,
i.e., the region $\theta\rightarrow0$ and $r\rightarrow r_{o}$.
Furthermore, at the ionospause nose, $dr/rd\theta$ can be approximated
by $dx/dy.$ The differential equation for the ionopause profile (Eq.
\ref{eq: dr/rdtheta}) at the vicinity of the ionopause nose can be
simplified to 

\begin{align}
\frac{dx}{dy} & =-\sqrt{\frac{P_{i}(r_{o})}{P_{i}(r)}-1}\nonumber \\
 & =-\sqrt{\frac{P_{i}(r_{o})}{P_{i}(r)}}\sqrt{1-\frac{P_{i}(r)}{P_{i}(r_{o})}}\nonumber \\
 & \simeq-\sqrt{1-\frac{P_{i}(r)}{P_{i}(r_{o})}}.\label{eq:dxdy}
\end{align}
Now we express $P_{i}(r)$ in Taylor series at $r=r_{o}$, then the
right hand side of the Eq. \ref{eq:dxdy} can be rewritten as
\begin{align}
-\sqrt{1-\frac{P_{i}(r)}{P_{i}(r_{o})}} & =-\sqrt{-(r-r_{o})\frac{P_{i}'(r_{o})}{P_{i}(r_{o})}-\frac{1}{2}(r-r_{o})^{2}\frac{P_{i}"(r_{o})}{P_{i}(r_{o})}-...}\nonumber \\
 & =-\sqrt{-(r-r_{o})\frac{P_{i}'(r_{o})}{P_{i}(r_{o})}}\sqrt{1+\frac{1}{2}(r-r_{o})\frac{P_{i}"(r_{o})}{P_{i}'(r_{o})}+...}\nonumber \\
 & \simeq-\sqrt{-(r-r_{o})\frac{P_{i}'(r_{o})}{P_{i}(r_{o})}}\left(1+\frac{1}{4}(r-r_{o})\frac{P_{i}"(r_{o})}{P_{i}'(r_{o})}+...\right).\label{eq:sqrt1-p}
\end{align}
Thus, now the differential equation for the ionopause profile near
the ionopause nose is 
\begin{equation}
\frac{dx}{dy}=-\sqrt{-(r-r_{o})\frac{P_{i}'(r_{o})}{P_{i}(r_{o})}}\left(1+\frac{1}{4}(r-r_{o})\frac{P_{i}"(r_{o})}{P_{i}'(r_{o})}+...\right)\label{eq:dxdyf}
\end{equation}

Also, by $x\rightarrow r_{o}$and $y\rightarrow0$ at the vicinity
of the ionopause nose, $r-r_{o}$ can be approximated by 

\begin{align}
r-r_{o} & =\sqrt{x^{2}+y^{2}}-r_{o}\nonumber \\
 & =\sqrt{\left[r_{o}+\left(x-r_{o}\right)\right]^{2}+y^{2}}-r_{o}\nonumber \\
 & =\sqrt{r_{o}^{2}+2\,r_{o}(x-r_{o})+(x-r_{o})^{2}+y^{2}}-r_{o}\\
 & =r_{o}\sqrt{1+2\,\frac{x-r_{o}}{r_{o}}+\frac{\left(x-r_{o}\right)^{2}}{r_{o}^{2}}+\frac{y^{2}}{r_{o}^{2}}}-r_{o}\nonumber \\
 & \simeq r_{o}(1+\frac{x-r_{o}}{r_{o}}+\frac{\left(x-r_{o}\right)^{2}}{2\,r_{o}^{2}}+\frac{y^{2}}{2\,r_{o}^{2}})-r_{o}\\
 & =x-r_{o}+\frac{\left(x-r_{o}\right)^{2}}{2\,r_{o}}+\frac{y^{2}}{2\,r_{o}}.\label{eq:r-ro-1}
\end{align}
By substituting the equation of the ionopause profile near the ionopause
nose (Eq.\ref{eq:expansion with Ro}) in to Eq.\ref{eq:r-ro-1}, we
get 
\begin{align}
r-r_{o} & =-\frac{1}{2R_{o}}y^{2}+\frac{\left(-\frac{1}{2R_{o}}y^{2}\right)^{2}}{2\,r_{o}}+\frac{y^{2}}{2\,r_{o}}\label{eq:r-ro2}\\
 & \simeq\frac{y}{2}^{2}\left(\frac{1}{r_{o}}-\frac{1}{R_{o}}\right)\label{eq:r-ro3}
\end{align}
In Eq.\ref{eq:r-ro2}, the second term at the right hand side is negligible
as $y\rightarrow0$ since it is of the order $y^{4}$ and the other
two terms are of the order $y^{2}$. 

We plug the $r-r_{o}$ relation (Eq. \ref{eq:r-ro3}) into the differential
equation of the ionopause profile (Eq. \ref{eq:dxdyf}), then integrate
the differential equation 
\begin{equation}
\sideset{}{_{r_{o}}^{x}}\int dx=\sideset{-}{_{0}^{y}}\int dy\,\sqrt{\frac{-1}{2}\frac{P_{i}'(r_{o})}{P_{i}(r_{o})}\left(\frac{1}{r_{o}}-\frac{1}{R_{o}}\right)}\left(y+\frac{1}{8}\left(\frac{1}{r_{o}}-\frac{1}{R_{o}}\right)\frac{P_{i}''(r_{o})}{P_{i}'(r_{o})}\,y^{3}+...\right).\label{eq:integration}
\end{equation}
Therefore we obtain the formula of the ionopause profile at the vicinity
of the ionopause:
\begin{equation}
x=r_{o}-\sqrt{-\frac{P_{i}'(r_{o})}{2P_{i}(r_{o})}\left(\frac{1}{r_{o}}-\frac{1}{R_{o}}\right)}\left(\frac{y^{2}}{2}+\frac{1}{32}\left(\frac{1}{r_{o}}-\frac{1}{R_{o}}\right)\frac{P_{i}''(r_{o})}{P_{i}'(r_{o})}\,y^{4}+...\right).\label{eq:ionopause_eq}
\end{equation}

By comparing Eq. \ref{eq:ionopause_eq} and Eq. \ref{eq:expansion with Ro},
finally, we get the equation of the radius of curvature at the ionopause
nose
\begin{equation}
R_{o}=\sqrt{-\frac{2P_{i}(r_{o})}{P_{i}'(r_{o})}\frac{R_{o}r_{o}}{R_{o}-r_{o}}}.\label{eq:exp of Ro}
\end{equation}
Eq.\ref{eq:exp of Ro} can be rearranged as a quadratic equation
\begin{equation}
R_{o}^{2}-r_{o}\,R_{o}-\sqrt{-\frac{2P_{i}(r_{o})}{P_{i}'(r_{o})}}\,r_{o}=0.\label{eq:quadraRo}
\end{equation}
Then we solve it and get
\begin{equation}
R_{o}=\frac{r_{o}\pm\sqrt{r_{o}^{2}+4\,\sqrt{-\frac{2P_{i}(r_{o})}{P_{i}'(r_{o})}}\,r_{o}}}{2}.\label{eq:Ro}
\end{equation}
We take the plus sign in the nominator since $R_{o}$ will be negative
if we take the minus sign, which is not physically allowable. Thus,
we obtain the expression of the radius of curvature at the ionopause
nose
\begin{equation}
R_{o}=\frac{r_{o}+\sqrt{r_{o}^{2}+4\,\sqrt{-\frac{2P_{i}(r_{o})}{P_{i}'(r_{o})}}\,r_{o}}}{2}.\label{eq:FinalRo}
\end{equation}

With the ionospheric pressure $P_{i}(r)$ of the form in Eq. \ref{eq:exp of pion-1},
the radius of curvature at the ionopause nose can be expressed as
\begin{equation}
R_{o}=\frac{r_{o}+\sqrt{r_{o}^{2}+8\,H\,r_{o}}}{2},\label{eq:radi_ionop}
\end{equation}
where $H$ is the scale height of the ionosphere. The expression of
the radius of curvature at the ionopause nose from our calculation
is the same as that from the Table A1 in Verigin \textit{et al.} (2003)
\cite{verigin2003planetary}. Thus, by plugging Eq.\ref{eq:radi_ionop}
into Eq.\ref{eq:expansion with Ro}, we can get the ionopause profile
near the ionopause nose 
\begin{equation}
x=r_{o}-\frac{1}{r_{o}+\sqrt{r_{o}^{2}+8\,H\,r_{o}}}\,y^{2}.\label{eq:ionopa_eq}
\end{equation}
In the derivation of the radius of curvature at the ionopause nose
and the ionopause profile near the ionopause nose, we made many assumptions
to get it. We have shown that the assumption we made above is all
valid by verifying our analytical results with the simulation results,
which will be given in Section \ref{subsec:Verification-of-the}.

This equation of the radius of the curvature at the ionopause (Eq.\ref{eq:radi_ionop})
tells that 
\begin{equation}
R_{o}\geq r_{o}.\label{eq:Ro>=00003Dro}
\end{equation}
The equality occurs as the ionospheric scale height $H$ is close
to zero. By the equation of the ionopause nose location (Eq.\ref{eq:exp of ro}),
we get that $r_{o}$ is equal to $r_{M,i}$ when $H$ is zero. Combining
the above relation, we found that 
\begin{equation}
R_{o}\rightarrow r_{M,i}\;{\rm as\;H\rightarrow0}.\label{eq:ro->rmi}
\end{equation}
This relation means that if the ionospheric pressure decreases very
sharply outward, the radius of the ionopause nose is about the distance
between the location of the ionospheric peak pressure and the planet
center. Furthermore, we can observe that $R_{o}$ is smaller as the
dynamic pressure of the solar wind is larger since $r_{o}$ is smaller.
These results are physically reasonable.

\subsection{Bow shock standoff distance $\Delta$\label{subsec:Bow-shock-standoff}}

\begin{figure}[H]
\begin{centering}
\includegraphics[width=0.7\columnwidth]{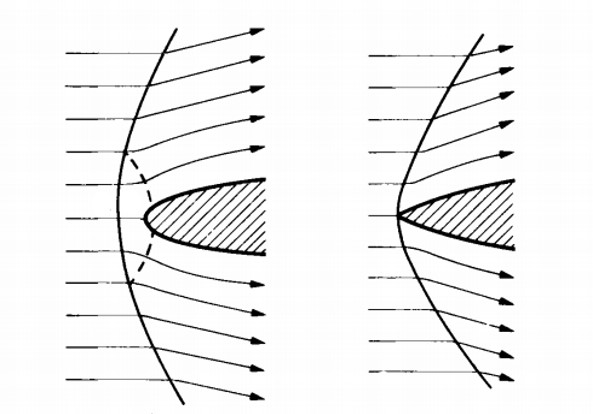}
\par\end{centering}
\caption{Schematic of the detached shock. Courtesy of \cite{landau1987fluid}\label{fig:detached} }

\end{figure}

The standoff distance of the bow shock $\Delta$ is determined by
the empirical model \cite{farris1994determining,spreiter1966hydromagnetic,verigin2003planetary,seiff1962recent}.
This empirical model is supported by gasdynamics experiment and observations
of the flow past the planets\cite{slavin1983solar}. The standoff
distance is expressed by the empirical model
\begin{equation}
\frac{\Delta}{R_{o}}=\alpha\frac{\rho_{\infty}}{\rho_{s}},\label{eq:standoff}
\end{equation}
where $\rho_{\infty}$and $\rho_{s}$ are the mass density before
and after shock, respectively and 
\[
\alpha\approx0.87
\]
 is the empirical coefficient from \cite{farris1994determining,verigin2003planetary}.
The bow shock nose is farther from the obstacle as the radius of curvature
of the obstacle nose is larger (Fig. \ref{fig:detached}). Bow shock
can touch the obstacle only when the leading end of the obstacle is
pointed. 

The density ratio across the shock is related to solar wind Mach number
and the specific heat ratio
\begin{equation}
\frac{\rho_{\infty}}{\rho_{s}}=\frac{(\gamma-1)M_{\infty}^{2}+2}{(\gamma+1)M_{\infty}^{2}}.\label{eq:ratio}
\end{equation}
Thus, in the condition of the solar wind Mach number much larger than
1, the standoff distance can be expressed as
\begin{equation}
\Delta=\alpha\ R_{o}\epsilon,\quad M_{\infty}\gg1,\label{eq:exp of Delta}
\end{equation}
where $\epsilon=\frac{\gamma-1}{\gamma+1}$. By substituting the Eq.
\ref{eq:radi_ionop} into Eq. \ref{eq:exp of Delta}, we get 
\begin{align}
\Delta & =\alpha\epsilon\frac{r_{o}+\sqrt{r_{o}^{2}+8\,H\,r_{o}}}{2},\quad M_{\infty}\gg1,\label{eq:deltaro}\\
 & =\frac{1}{2}\,r_{o}\,\alpha\,\epsilon\left(1+\sqrt{1+8\,H/r_{o}}\right),\quad M_{\infty}\gg1.\nonumber 
\end{align}
According to this relation, we can obtain the value of the standoff
distance if we have the ratio of $H$ (scale height of the ionosphere)
to $r_{o}$ (the distance between ionopause and the center of the
planet).

\subsection{The formula of the bow shock nose location\label{subsec:shock location}}

Thus, combining the results above, the shock nose location can be
written as 

\begin{align}
r_{s} & =r_{o}+\Delta,\label{eq: shock nose location}\\
 & =r_{o}+\frac{1}{2}\,\alpha\,\epsilon\left(r_{o}+\sqrt{r_{o}^{2}+8\,H\,r_{o}}\right),\quad M_{\infty}\gg1,\nonumber 
\end{align}
 where 
\begin{equation}
r_{o}=r_{M,i}+H\ {\rm ln}(\frac{P_{M,i}}{k\rho_{\infty}v_{\infty}^{2}}),\label{eq:oexp}
\end{equation}
, scale height $H=k_{B}T/(mg)$ and $\epsilon=\frac{\gamma-1}{\gamma+1}$.

Note that this equation is only valid for the sonic Mach number of
the solar wind much larger than 1. As we can see in the shock nose
equation (Eq. \ref{eq: shock nose location}): the shorter the scale
height $H$ or the larger the dynamic pressure of the solar wind $\rho_{\infty}v_{\infty}^{2}$,
bow shock nose is closer to the planet. This result is reasonable
and intuitive.

\pagebreak{}

\section{Comparison with the numerical simulation and spacecraft measurement
results\label{subsec:Comparison-theory}}

The comparison of our formula of the bow shock location with the numerical
simulation and the spacecraft measurement results is given in this
section.

\subsection{Verification of the analytical form of the radius of curvature by
simulation\label{subsec:Verification-of-the}}

In this subsection, we verify the analytical results of the radius
of curvature from the Section \ref{subsec:Radius-of-curvature} by
the numerical simulation. In our calculation, we analytically solve
the differential equation of the ionopause profile (Eq. \ref{eq: dr/rdtheta})
near the ionopause nose ($\theta\rightarrow0$) with the initial condition
$r\left(\theta=0\right)=r_{o}$ to the get the equation of the ionopause
profile near the ionopause nose (Eq. \ref{eq:ionopa_eq}). The radius
of curvature at the ionopause nose is also obtained in Eq. \ref{eq:radi_ionop}.
In the derivation of the equation of the ionopause profile near the
ionopause nose and the radius of curvature at the ionopause nose,
we made many assumptions, so we will verify that the assumptions are
valid by numerical simulation.

\begin{figure}[H]
\begin{centering}
\includegraphics[width=0.8\columnwidth]{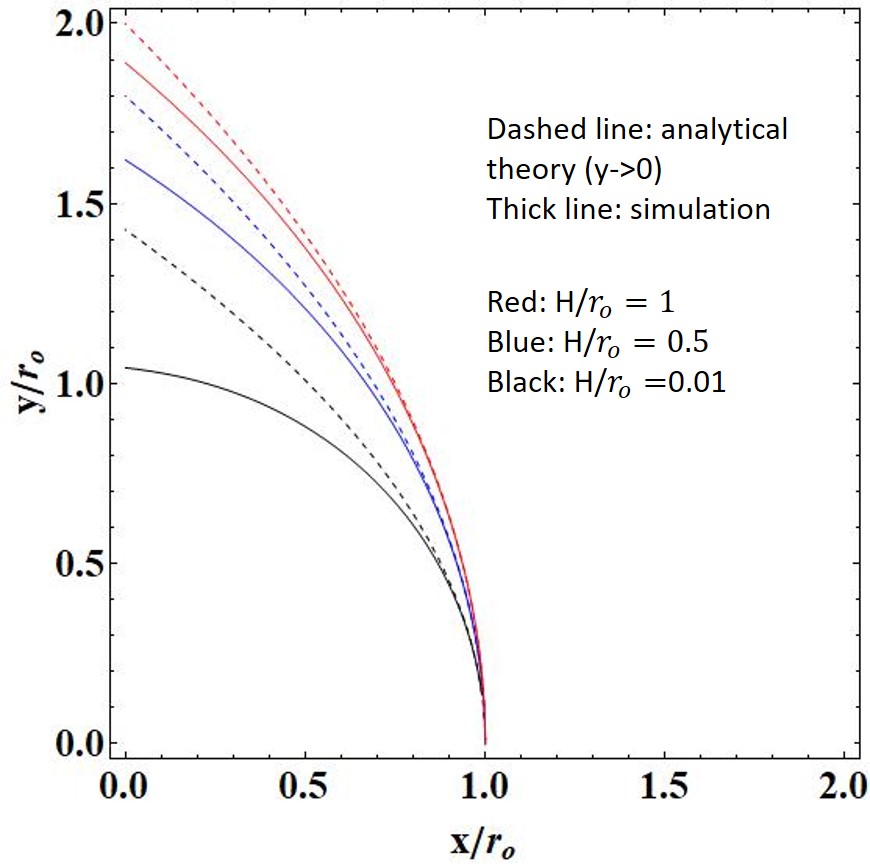}
\par\end{centering}
\caption{Ionopause profile calculated from analytical theory (Eq. \ref{eq:ionopa_eq})
and numerical simulation. The analytical results is calculated under
the approximation that $\theta\rightarrow0$ in polar coordinate or
$y\rightarrow0$ in Cartesian coordinate. \label{fig:Ionopause-profile-calculated}}
\end{figure}

We use the function \textit{NDSolve} in the software \textit{Mathematica
}to solve the differential equation of the ionopause profile (Eq.
\ref{eq: dr/rdtheta}) to get the ionopause profile $r=r(\theta)$.
We first compare the numerical result with our analytical result of
the ionopause profile near the ionopause nose (Eq. \ref{eq:ionopa_eq}).
Fig. \ref{fig:Ionopause-profile-calculated} shows the comparison
of the ionopause profile from analytical theory and numerical simulation.
We can observe that, from both the simulation and analytical results,
the ionopause follows the rule that 
\begin{equation}
r\geq r_{o},\label{eq:rlargerro-1}
\end{equation}
 which we show in the section \ref{subsec:Radius-of-curvature}. Also,
we can see that the ionopause profile from analytical theory and numerical
simulation match well near the nose position ($y\rightarrow0$), which
is reasonable since the analytical result is calculated under the
approximation that $\theta\rightarrow0$ in polar coordinate or $y\rightarrow0$
in Cartesian coordinate. 

\begin{figure}[H]
\begin{centering}
\includegraphics[width=0.8\columnwidth]{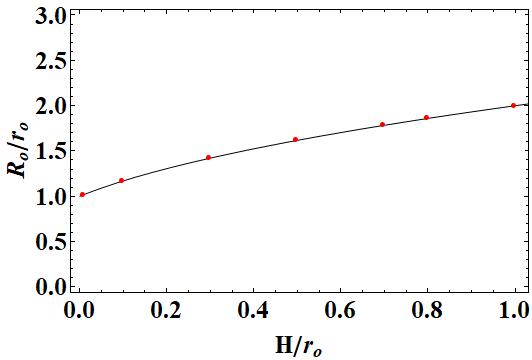}
\par\end{centering}
\caption{Radius of curvature at the ionopause nose from analytical theory (Eq.
\ref{eq:radi_ionop}) and numerical simulation. The thick line is
the analytical theory (Eq. \ref{eq:radi_ionop}) and the red dots
are simulation results. The cases with $H/r_{o}=0.01,\,0.1,\,0.3,\,0.5,\,0.7,\,0.8,\,1.0$
are considered. \label{fig:radius-of-curvature}}
\end{figure}

We then compare the numerical results with our analytical result of
the radius of curvature at the ionopause nose (Eq. \ref{eq:radi_ionop}).
In the numerical simulation, we numerically solve the differential
equation of the ionopause profile (Eq. \ref{eq: dr/rdtheta}) to get
the ionopause profile $r=r(\theta)$ using the function \textit{NDSolve}
in \textit{Mathematica} and then calculate the radius of curvature
at $\theta=0$ in polar coordinate using the formula
\begin{equation}
R(\theta)=\frac{\left(r^{2}(\theta)+r'^{2}(\theta)\right)^{3/2}}{|r^{2}(\theta)+2\,r'^{2}(\theta)-r(\theta)r''(\theta)|}.\label{eq:radi_polar}
\end{equation}
The comparison results are shown in Fig. \ref{fig:radius-of-curvature}.
We can see that the difference between analytical and numerical results
is smaller than 1 percent. Thus, our analytical form of the radius
of curvature at the ionopause nose (Eq. \ref{eq:radi_ionop}) is verified.
And we can say that the assumptions we made in the derivation are
valid.

\subsection{Comparison with hydrodynamics simulation}

\begin{figure}[H]
\begin{centering}
\includegraphics[width=0.7\columnwidth]{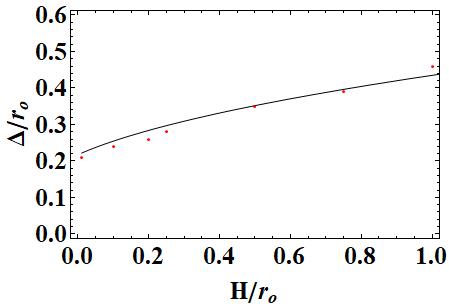}
\par\end{centering}
\caption{Comparison of the derived standoff distance formula (Eq. \ref{eq:sta_dt})
with the gasdynamics simulation results from \textit{Spreiter et al.},
1970\cite{spreiter1970solar}. The red dots are the simulation results;
the black line is Eq. \ref{eq:sta_dt}. The cases with $H/r_{o}=0.01,\,0.1,\,0.2,\,0.25,\,0.5,\,0.75,\,1.0$
are considered.\label{fig:Comparison-of-the}}

\end{figure}

We now want to compare our formula of the bow shock nose with the
simulation and spacecraft measurement results. Our derived formula
of the bow shock location is $r_{s}=r_{o}+\Delta$ and the standoff
distance is 
\begin{equation}
\frac{\Delta}{r_{o}}=\frac{1}{2}\,\alpha\,\epsilon\left(1+\sqrt{1+8\,H/r_{o}}\right),\quad M_{\infty}\gg1,\label{eq:sta_dt}
\end{equation}
where $\epsilon=\frac{\gamma-1}{\gamma+1}$. In Fig. \ref{fig:Comparison-of-the},
we compare our formula of the standoff distance with the nonlinear
gasdynamics simulation result for the bow shock profile, which is
referred to the paper by Spreiter \textit{et al.}, 1970\cite{spreiter1970solar}.

 As we can observe in the comparison, our derived formula (Eq. \ref{eq:sta_dt})
and the simulation results match well and both show that the standoff
distance becomes larger with the increasing scale heights of the ionosphere.
We can conclude that the theory is validated by the simulation results.
Note that this formula of the bow shock standoff distance is only
applied for the solar wind interacting with the unmangetized planet
and the Mach number of the solar wind solar wind must be much larger
than 1. The other assumption in this theory is that the ionosphere
of the unmagnetized planet is in hydrostatic condition, resulting
in the thermal pressure exponentially decaying outward. 

\subsection{Comparison with spacecraft measurements in Martian bow shock}

We investigate the influence of the shock location from solar parameters,
which control the $P_{dyn}$ (dynamic pressure of the solar wind),
$P_{M,i}$ (peak pressure of the ionosphere), $H=k_{B}T/mg$ (scale
height of the ionosphere). Our derived formula of the shock nose location
is 
\begin{equation}
r_{s}=r_{o}+\frac{1}{2}\,r_{o}\,\alpha\,\epsilon\left(1+\sqrt{1+8\,H/r_{o}}\right),\quad M_{\infty}\gg1,\label{eq:rs}
\end{equation}
where $\epsilon=\frac{\gamma-1}{\gamma+1}$ and
\begin{equation}
r_{o}=r_{M,i}+H\ {\rm ln}(\frac{P_{M,i}}{k\,P_{dyn}}).\label{eq:oexp-1}
\end{equation}

According to the spacecraft measurement\cite{vogt2015ionopause} (Fig.
\ref{fig:The-response-of}), the measurement data shows that the Martian
bow shock location increases linearly with the increasing EUV flux,
i.e.,
\begin{equation}
r_{s}\propto0.11\,l_{EUV},\label{eq:rspropleuv}
\end{equation}
 but it reduces through a power law relationship with solar wind dynamic
pressure, i.e.,
\begin{equation}
r_{s}\propto p_{dyn}^{-0.02}.\label{eq:obser}
\end{equation}
We can observe that our relation in Eq. \ref{eq:rs} and the spacecraft
measurement results in Eq. \ref{eq:obser} both shows that the increasing
solar EUV flux and decreasing solar wind dynamic pressure will increase
the bow shock location. The increasing solar EUV flux will cause the
$P_{M,i}$ (peak pressure of the ionosphere) increase via increasing
the ionization rate. Furthermore, the increasing solar EUV flux will
let the temperature increase, i.e., a larger scale height $H$. Since
the dynamics pressure $P_{dyn}$ term is in the logarithm in our relation
in Eq. \ref{eq:rs}, we suggest that the variation of dynamics pressure
has less impact on the shock location than the EUV flux, which controls
the scale height $H$ and ionospheric peak pressure $P_{M,i}$. Thus,
our derived formula is qualitatively consistent with the spacecraft
measurement results that the shock nose location is more sensitive
to the solar EUV flux than solar wind dynamics pressure.

The power-law dependence of Martian bow shock location on solar wind
dynamics pressure according to spacecraft measurement results (Eq.
\ref{eq:obser}) can be rewritten 
\begin{equation}
\frac{d\,r_{s}}{r_{s}}=-C\,\frac{d\,P_{dyn}}{P_{dyn}}.\label{eq:Pdyn-rs}
\end{equation}
where 
\[
C=0.02.
\]

Now we derive the mathematical expression of the $C$ using our formula
of the shock nose location (Eq. \ref{eq:rs}) and then compare it
with the spacecraft measurement results. In the Martian condition,
the scale height $H$ is much shorter than the distance between the
Martian ionopause nose and the Mars center $r_{o}$, so the square
root term in our formula of the bow shock location (Eq. \ref{eq:rs})
can be expanded, then it can be reduced to

\begin{align}
r_{s} & \simeq r_{o}+\frac{1}{2}\,r_{o}\,\alpha\,\epsilon\,(1+1+4\frac{H}{r_{o}}),\label{eq:rsexpan1}\\
 & =r_{o}(1+\alpha\,\epsilon)+2\,\alpha\,\epsilon\,H,\label{eq:rsexpan2}\\
 & =(1+\alpha\,\epsilon)\,r_{M,i}+(1+\alpha\,\epsilon)\,H\,{\rm ln}\left(\frac{P_{M,i}}{k\,P_{dyn}}\right)+2\,\alpha\,\epsilon\,H.\label{eq:rs-expan}
\end{align}
We are interested in the bow shock location $r_{s}$dependence on
$H,\;P_{M,i},\;P_{dyn}$. The total variation of the bow shock location
$r_{s}$ is written as

\begin{equation}
d\,r_{s}=\left[\left(1+\alpha\,\epsilon\right){\rm ln}\left(\frac{P_{M,i}}{k\,P_{dyn}}\right)+2\,\alpha\,\epsilon\right]\,d\,H+\left[\left(1+\alpha\,\epsilon\right)\,\frac{H}{P_{M,i}}\right]\,d\,P_{M,i}+\left[-\left(1+\alpha\,\epsilon\right)\,\frac{H}{P_{dyn}}\right]\,d\,P_{dyn}.\label{eq:drs}
\end{equation}
The bow shock location $r_{s}$ dependence on the solar wind dynamics
pressure $P_{dyn}$ is 

\begin{equation}
\frac{d\,r_{s}}{r_{s}}=-(1+\alpha\,\epsilon)\,\frac{H}{r_{s}}\,\frac{d\,P_{dyn}}{P_{dyn}},\label{eq:drs/rs}
\end{equation}
where we assume the $H\;{\rm and}\;P_{M,i}$ are fixed. Thus, we get
the expression of $C$ as 
\begin{equation}
C=(1+\alpha\,\epsilon)\,\frac{H}{r_{s}}.\label{eq:C}
\end{equation}
$C$ is approximately the ratio between ionospheric scale length and
the distance between bow shock nose location and the planet center.
According to spacecraft measurement results in the Martian environment
\cite{vogt2015ionopause}, the ionospheric scale height $H$ is about
100 km and the distance between the Martian bow shock location between
the Mars center is about 2.5 Mars radius (8473 km). Also, with $\alpha=0.87$
and $\epsilon=0.25$, the value of $C$ calculated from Eq. \ref{eq:C}
is 0.014, which is at the same order of the results from spacecraft
measurement results ($C=0.02$). Thus, in terms of the bow shock location
dependence on the solar wind dynamics pressure, our formula is verified
by the spacecraft measurement results. This formula can be used for
the future spacecraft measurement prediction.

\pagebreak{}

\section{Conclusion\label{sec:Conclusion}}

It is not well understood how the solar parameters, such as solar
wind dynamics pressure, solar wind EUV flux and ionospheric pressure,
influence the location of the Martian (unmagnetized planet) bow shock.
The location of the bow shock produced from the solar wind interacting
with the unmagnetized planet has been theoretically investigated.
The formula for the location of the bow shock produced from the solar
wind interacting with the unmagnetized planet is presented. The bow
shock location is the sum of the ionopause location and standoff distance.
The whole calculation is based on the gasdynamics formulation since
the magnetic effect can be neglected in space environment for our
purpose. We determine the ionopause nose location using thermal pressure
continuity according to tangential discontinuity. The standoff distance
of the bow shock produced by the supersonic plasma jet with sonic
Mach number much larger than 1 past the obstacle is given in the empirical
formula that the standoff distance is proportional to the radius of
curvature at obstacle leading end. The formula of the shock nose position
was derived and showed that the shock nose location increases with
the increasing scale height of ionosphere, the decreasing dynamic
pressure of the solar wind and the increasing peak pressure of the
ionosphere. Furthermore, we derived the equation of the ionopause
profile around the nose. Our derived theory is consistent with the
numerical simulation. The derived analytical form of the radius of
curvature at the obstacle leading end is verified by the numerical
simulation. Also, the derived formula of the bow shock location for
the unmagnetized planet is consistent with the results from the gasdynamics
simulations. 

Furthermore, we derived that the relation between the unmagnetized
planet bow shock location and solar wind dynamics pressure, which
is $r_{s}\propto p_{dyn}^{-C},$ where the constant $C$ is roughly
the ratio between ionospheric scale length and the distance between
bow shock nose location and the The constant $C$ calculated from
the Martian parameter with our formula is at the same order as the
results from the spacecraft measurements by Mars Express\cite{hall2016annual}.
Also, our derived formula is qualitatively consistent with the spacecraft
measurement results that the shock nose location is more sensitive
to the solar EUV flux than solar wind dynamics pressure.

Since our model only provide the relation between bow shock location
and the solar wind dynamics pressure, in order to have a more thorough
comparison of our theory with the measurement results, we have to
study how the solar EUV flux controls the ionospheric pressure, which
is related to the photoionization from EUV radiation. On the other
hand, throughout this work, we neglect the effect of the magnetic
field for simplicity. In fact, Mars has some local magnetic field,
which can influence the bow shock nose location. The magnetic field
also influences the formation mechanisms of the bow shock and the
ionopause. In the realistic condition, the interaction between the
interplanetary magnetic field and the ionosphere will generate the
``induced magnetosphere'' and ``magnetic pile-up boundary''. Furthermore,
the magnetic draping effect will occur. Although the detailed study
of how important the interplanetary magnetic field plays the role
on controlling the bow shock location is yet to be done, our model
with no magnetic field effect can now accurately predict the bow shock
nose location and its relationship with the solar wind dynamics pressure,
which has been verified by both numerical simulation and spacecraft
measurement results. 

\pagebreak{}

\section{Acknowledgment}

I.L.Yeh acknowledges support through MOST grant No. and the helpful
discussion with Frank C.Z. Cheng and Ling-Hsiao Lyu. I.L. Yeh did
all the calculation and simulation and wrote the paper draft. P.Y.
Chang came up with this project and reviewed the draft. Sunny W.Y.
Tam reviewed the draft and provided the insights on the comparison
with the theory and spacecraft measurement results. 

\pagebreak{}\bibliographystyle{unsrt}
\addcontentsline{toc}{section}{\refname}\bibliography{bs_ref}

\end{CJK}

\end{document}